\documentclass[doublecol]{epl2}

\usepackage{amssymb}

\title{Phase Transition of $XY$ Model in Heptagonal Lattice}
\shorttitle{Phase Transition of $XY$ Model in Heptagonal Lattice}
\author{Seung Ki Baek\inst{1}, Petter Minnhagen\inst{2} \and Beom Jun
Kim\inst{1}}
\shortauthor{S.K. Baek \etal}
\institute{
	\inst{1} Department of Physics, BK21 Physics Research Division, and
Institute of Basic Science, Sungkyunkwan University, Suwon 440-746,
Republic of Korea\\
	\inst{2} Department of Theoretical Physics, Ume{\aa} University, 901 87
Ume{\aa}, Sweden
}


\abstract{
We numerically investigate the nature of the phase transition of the $XY$ model
in the heptagonal lattice with the negative curvature, in comparison
to other interaction structures such as a flat two-dimensional (2D)
square lattice and a small-world network. Although the heptagonal lattice
has a very short characteristic path length like the small-world network
structure, it is revealed via calculation of the Binder's cumulant
that the former exhibits a zero-temperature phase transition while the 
latter has the finite-temperature transition of the mean-field nature. 
Through the computation of the vortex density as well as
the correlation function in the low-temperature
approximation, we show that the absence of the phase transition originates
from the strong spinwave-type fluctuation, which is discussed in relation
to the usual 2D $XY$ model.  
}

\begin{document}

\maketitle

A common belief in the complex network researches is that 
critical phenomena in networks often exhibit mean-field
natures. 
For the small-world network structure proposed by Watts and 
Strogatz (WS)~\cite{WS}, it is well-known that statistical mechanical 
model systems such as the Ising and the $XY$ models have 
phase transitions completely different from their counterparts in 
one- and two-dimensional (2D) regular lattices~\cite{smallIsing,smallxy}. 
Furthermore, the universality class of the phase transition has 
been known to undergo a dramatic change as soon as the density
of shortcuts has a finite nonzero value, which is accompanied
by the structural phase transition from the large world to 
the small world. 
These existing findings imply the importance
of the topological interaction structure in the determination
of the universality class of the phase transition.
Recently, the Ising model in the heptagonal lattice, which has complicated
geometry with a negative curvature~\cite{negative}, has been shown to possess
the mean-field critical exponents, consistently with the common
belief~\cite{Shima1,Shima2}.

\begin{figure}
\begin{center}
\includegraphics[width=0.35\textwidth]{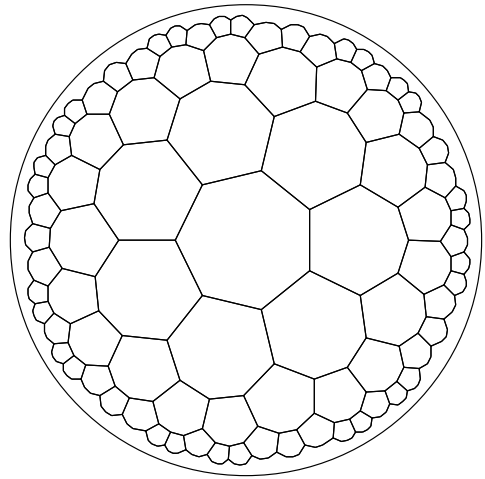}
\caption{
Schematic representation of a heptagonal lattice upto the 4th level ($L=4$), 
projected on the Poincar\'{e} disk~\cite{book}. The surrounding circle corresponds to 
the points at infinity, and the lattice within it is composed of congruent 
heptagons with respect to the negative curvature metric~\cite{coxeter}.
}
\label{fig:heptagon}
\end{center}
\end{figure}

In the present Letter, we take  the one of the mostly well-studied
statistical mechanical model systems, the $XY$ model, and study the nature of
the phase transition in the heptagonal lattice. The comparison with the usual
flat 2D  square lattice sheds a light on the role of the geometry
with different curvatures, while the comparison with the small-world network
structure indicates that, different from a naive expectation,
the short path length cannot solely determine the
universality class. Extensive Monte-Carlo computations of the Binder's
cumulant  and the vortex density, combined with the numerical 
calculation of the spin-spin correlation function, bring us to the conclusion
that the absence of the finite-temperature phase transition in the heptagonal
lattice originates not from the vortex fluctuation, but from the strong
spinwave fluctuation.  
 
\begin{figure}
\includegraphics[width=0.48\textwidth]{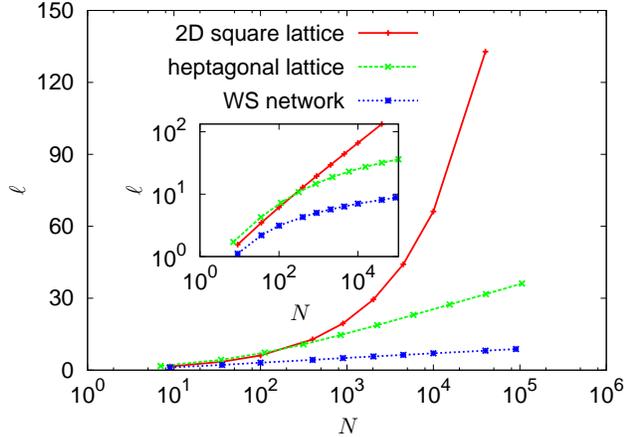}
\caption{(Color online) The characteristic path length $\ell$ versus
the system size $N$ for the 2D square lattice, heptagonal lattice, and the WS
network (from top to bottom). Both the heptagonal lattice
and the WS network exhibit $\ell \sim \ln N$, while the
2D square lattice shows $\ell \sim N^{1/2}$. The latter is
more clearly seen in the inset in which the same data points
are plotted in log-log scales.}
\label{fig:l}
\end{figure}

We start from a brief description of the heptagonal 
lattice~\cite{Shima1,Shima2} (see Fig.~\ref{fig:heptagon}).
Suppose that we start constructing the heptagonal lattice from 
a single heptagon (the 1st level), the central one in Fig.~\ref{fig:heptagon}. 
Its seven nearest neighbor heptagons constitute the 2nd level,
and the next nearest neighbors the 3rd level, and so on.
We denote the set of heptagons in the $l$th level as $H(l)$ 
and call set of vertices on the outward boundary of $H(l)$
as the $l$th layer. The total number $N(L)$ of spins
in the heptagonal lattice containing all the layers upto $L$th one 
has been shown to increase exponentially with $L$~\cite{Shima1,Shima2}, 
which implies the following two interesting features:
(i) The surface-volume ratio does not vanish 
but remains finite in the thermodynamic limit.
(ii) The average path length $\ell$ increases logarithmically 
with the system size $N$, i.e., $\ell \sim \ln N$, since
the spins separated by a long distance on a given layer are
connected by much shorter path through inter-layer connections.
In Fig.~\ref{fig:l} we plot $\ell$ versus $N$ for the heptagonal lattice,
the 2D square lattice, and the WS small-world
network. It is clearly seen that  $\ell \sim \ln N$ 
for the heptagonal lattice and for the WS network,
and $\ell \sim \sqrt{N}$ for the 2D square lattice.
From the above comparisons, one may naively expect the same 
universality class both for the heptagonal lattice and the WS 
network, however, it is clearly shown below that this is not the case.


The $XY$ model is one of the most important model systems 
not only in statistical physics but also in condensed-matter physics
due to the existences of real-world examples, 
and is described by the Hamiltonian~\cite{kosterlitz}:
\begin{equation}
{\cal H} = -J \sum_{\langle ij \rangle} \cos(\phi_i - \phi_j),
\end{equation}
where $J$ is the coupling constant, the sum is over all
nearest pairs in the system, and the variable $\phi_i$ at
the $i$th site can represent either the direction of
the two-dimensional spin in magnetic systems or the phase
of the complex order parameter in superconducting 
systems~\cite{rmp:minnhagen}. 
In a two-dimensional lattice, the Mermin-Wagner theorem~\cite{mermin} 
states that there cannot be a true long-range order at finite temperatures, 
however, it has been known that the 2D $XY$ model exhibit a quasi-long-range
order at low temperatures, characterized by the algebraic decay of
the correlation function.

\begin{figure}
\includegraphics[width=.48\textwidth]{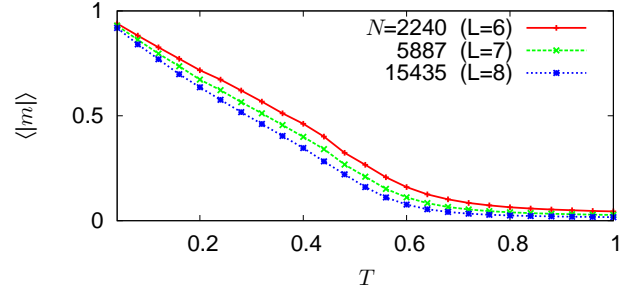}
\caption{(Color online) Ferromagnetic order parameter $m$ of heptagonal
lattices, as a function of the temperature $T$ in units of $J/k_B$. The
spins in the two outmost layers are excluded in computing $m$ to remove the
undesirable boundary effects.}
\label{fig:magnet}
\end{figure}

\begin{figure}
\includegraphics[width=.48\textwidth]{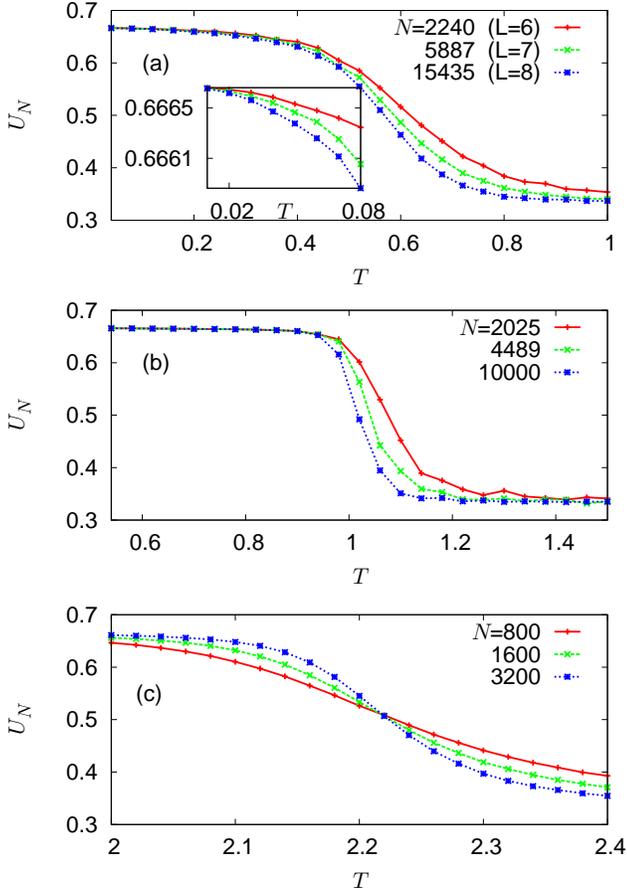}
\caption{(Color online) Binder's cumulant $U_N$ versus the temperature $T$
for (a) heptagonal lattices, (b) 2D regular square lattices,  
and (c) WS networks 
(Data in Ref.~\cite{smallxy} are used). 
In (a), we have excluded  two outmost layers in the computation 
of $U_N$ to avoid surface effects. 
Inset of (a): expansion of the temperature range
$T < 0.1$, showing that $U_N$'s neither cross nor merge.}
\label{fig:binder}
\end{figure}

We perform Monte Carlo simulations based on the Metropolis algorithm.
The ferromagnetic order parameter 
$m \equiv (1/N)\sum_j e^{i\phi_j}$ is computed and the Binder's cumulant 
for the system of the size $N$
\begin{equation}
U_N = 1-\frac{\langle m^4 \rangle }{3\langle m^2\rangle^2} 
\end{equation}
is measured with the ensemble average $\langle \cdots \rangle$.
We plot $\langle |m| \rangle$ in Fig.~\ref{fig:magnet} as a function of the temperature $T$ in
units of $J/k_B$ with the Boltzmann constant $k_B$, which clearly indicates
that the heptagonal lattice lacks a true long-range order, qualitatively
similarly to the 2D regular square lattice. 
The lowest excitation in the $XY$ model is the gapless spinwave
excitations, and thus there is no essential singularity at zero
temperature in thermodynamic quantities like susceptibility.
The next question for the
presence of a quasi-long-range order can be answered by observing the
Binder's cumulant, $U_N$.
Figure~\ref{fig:binder} shows $U_N$ versus the temperature $T$
for (a) the heptagonal lattice, (b) the 2D square lattice, 
and (c) the WS small-world network. As were already found in
the literature, the 2D square lattice shows merging of $U_N$ for different $N$'s
in the whole low-temperature phase with the quasi-long-range 
order~\cite{Loison}, while the WS network exhibits a
unique crossing at a well-defined critical temperature 
splitting the truly ordered phase and disordered phase~\cite{smallxy}.
The merging of the Binder's cumulant for the 2D square lattice
is due to the diverging correlation length in the whole low-temperature
phase, which has also been detected by the merging of the ratios of 
two correlation functions~\cite{okabe}.

For the heptagonal lattices [see Fig.~\ref{fig:binder}(a)], 
we exclude spins in outmost two layers in the computation of $m$
for a given system size in order
to avoid the artifact originating from the nonvanishing surface-volume
ratio in the thermodynamic limit. The Binder's cumulant for
heptagonal lattices in Fig.~\ref{fig:binder}(a) displays a completely different
behavior than for 2D regular lattice and WS networks: 
The curves for $U_N$ obtained from different sizes do not
merge even at very low temperatures $T<0.1$ [see Inset of Fig.~\ref{fig:binder}(a)]. 
Consequently, we conclude that
the $XY$ model in the heptagonal lattice is disordered at any
finite non-zero temperature and does not even possess the quasi-long-range
order present in the 2D regular lattice. This is a remarkably surprising
result from the viewpoint of the topological interaction structure:
In contrast to the WS network, the short path length in the heptagonal 
lattice does not increase the effective dimension of the system 
making the ordered state more probable. We next pursue the answer to
the question on the origin of the enhanced fluctuation in the heptagonal
lattice. 

\begin{figure}
\includegraphics[width=.48\textwidth]{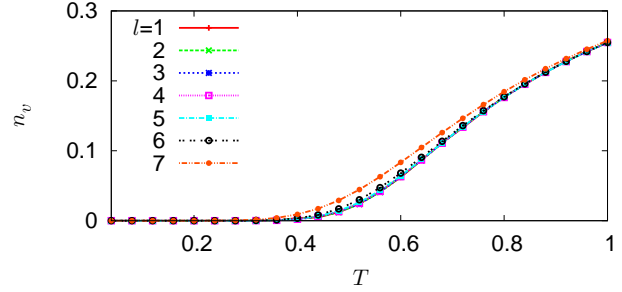}
\caption{(Color online) Vortex number density $n_v$ in a heptagonal lattice 
with seven layers as a function of $T$ is measured for each
level $l=1,2,\cdots, 7$. The proliferation of vortices occurs around
$T=0.4$, indicating that the absence of order in Fig.~\ref{fig:binder}(a)
is not from the vortex fluctuation.}
\label{fig:vortex}
\end{figure}

The quasi-long-range order present in the 2D $XY$ model is known to be broken 
by the proliferation of free vortices~\cite{kosterlitz}. Accordingly, 
the above found absence of the quasi-long-range order
in the heptagonal $XY$ model at finite temperatures could be explained
by the strong vortex fluctuation even at extremely low temperatures.
In order to test this possible scenario, we measure
the vortex number density $n_v(l)$  for the $l$th level heptagons  defined as 
\begin{equation}
n_v(l) \equiv \frac{1}{2\pi} \left\langle \frac{1}{| H(l) |}\sum_{p \in H(l)}  
   \left|{\sum}^p (\phi_i-\phi_j)\right|\right\rangle, 
\end{equation}
where $| H(l) |$ is the number of heptagons in the $l$th level,
the summation ${\sum}^p$ is taken in the counter-clockwise 
direction over the bonds surrounding the heptagon $p$, 
and the phase difference $\phi_i -\phi_j$ is defined modulo $2\pi$.
Although $n_v$ cannot detect the difference between the free
and bound vortices, it reflects the Kosterlitz-Thouless (KT)
transition since it changes from zero to nonzero smoothly around the 
transition temperature in the 2D $XY$ model.
We in Fig.~\ref{fig:vortex} plot $n_v$ as a function of $T$
for the heptagonal $XY$ model composed of seven layers ($L=7$): 
Vortices are observed to be generated at some finite temperature
around $T=0.4$ and the number density for each layer does not
show a significant difference. Overall, one sees that the temperature
region around $T=0.4$ in which thermal vortices are generated is
well above the zero temperature, which leads us to the conclusion
that the absence of any long-range order (be it genuine or quasi)
found in the heptagonal $XY$ model cannot be explained from 
the vortex degree of freedom. Accordingly, we below focus
on the spinwave excitation in the $XY$ model in the low-temperature
regime.

The spin-spin correlation function 
within the low-temperature approximation, i.e., 
$\cos\theta \approx 1 - \theta^2/2$, is
easily computed~\cite{izyumov} to yield
\begin{eqnarray}
\label{eq:C}
C_{jk} &\equiv& \langle  e^{i(\phi_j - \phi_k)}\rangle
\approx  e^{-g_{jk}/T},   \\
\label{eq:g}
g_{jk} &\equiv&  (1/2)(G_{jj} - G_{jk} + G_{kk} - G_{kj}), 
\end{eqnarray}
where the lattice Greens function $G_{jk}$ is the inverse
of the discrete Laplacian~\cite{neton} defined by 
$G_{jk}^{-1} \equiv k_j \delta_{jk} - A_{jk}$  
with the degree $k_j$ and the adjacency matrix $A_{jk}$~\cite{kahng}.
It is natural to 
measure the distance between the two spins at the $j$th and 
the $k$th vertices by the shortest path length, which is especially
important for interaction structures not put on geography, like
the WS network. 
If there exist paths of different lengths, we believe that
the path with the shortest length must give the most contribution
to the correlation function.

\begin{figure}
\includegraphics[width=.48\textwidth]{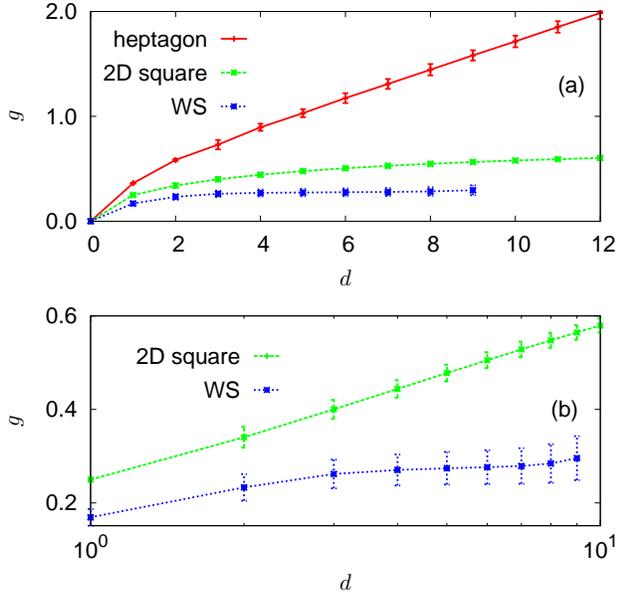}
\caption{(Color online) The low-temperature
correlation function is shown in the form of $g(d)$,
which is defined as the average of $g_{jk}$ for all pairs of $(j,k)$'s
separated by the given shortest path length $d$ [see Eqs.~(\ref{eq:C})  
and (\ref{eq:g})].
In (a), $g$ increases linearly with $d$ for the heptagonal lattice
(with $L=6$), 
indicating the exponential decay of the correlation function.  
In (b) the same data are plotted in lin-log scales, and
$g$ increases logarithmically for the 2D square lattice ($N = 40 \times 40$). 
It is seen that for the WS network ($N=1600$)
$g(d)$ saturates toward a finite value as $d$ is increased, indicating
the existence of the ordered phase at low temperatures.
}
\label{fig:green}
\end{figure}

In Fig.~\ref{fig:green}, we present spatial average $g(d)$
of $g_{jk}$ in Eq.~(\ref{eq:g}) taken for all pairs split
by the shortest path length $d$. 
The logarithmic increase of $g(d)$, 
and accordingly the algebraic decay of $C(d)$ in Eq.~(\ref{eq:C}),
for the 2D square lattice clearly shown in Fig.~\ref{fig:green}(b)
indicates  the existence of the quasi-long-range order in the
low-temperature phase below the KT transition~\cite{kosterlitz}. 
In Fig.~\ref{fig:green}
it should be noted that the absence of the long-range order
is reflected as the divergence of  $g(d)$ as $d$ is increased.
For the WS network, on the other hand, $g(d)$ does not 
diverge, but saturates to a well-defined value, which is in an agreement with
the expectation from Ref.~\cite{smallxy}: The existence of the
true long-range order of the $XY$ model in the WS network at
sufficiently low temperatures should be
reflected as a finite and nonvanishing correlation function $C_{jk}$
as $d_{jk}$ is increased.
In contrast, the behavior of $g(d)$ for the heptagonal lattice
in Fig.~\ref{fig:green}(a) is completely different both from
the 2D square lattice and the WS network: $g(d)$ increases
with $d$ like the 2D square lattice [shown in Fig.~\ref{fig:green}(b)],
however, not logarithmically but {\em linearly}. The linear increase
of $g$ with $d$ implies that the correlation function $C(d)$ 
decays {\em exponentially}, which clearly implies the absence 
of any long-range order at sufficiently low temperatures.
As a consequence, we conclude that the absence of the ordered phase
in the heptagonal $XY$ model originates not from the vortex fluctuation
but from the spinwave excitation, which is strong enough to destroy
even the quasi-long-range order present in the low-temperature
KT phase of the 2D square $XY$ model.

We emphasize that our observation of the absence of the ordered phase
in the heptagonal $XY$ model is contrasted to the report 
for the heptagonal Ising model~\cite{Shima1,Shima2}: 
The geometrical difference between the planar 2D lattice and
the heptagonal lattice with a negative curvature only changes 
the critical exponents of the Ising model to the mean-field ones. 
In other words, for the Ising model the overall strength of fluctuation 
appears to be reduced by an increase of an effective dimensionality,
reflected as the small-world behavior of very short path lengths, 
similarly to the WS networks. In a sharp contrast, 
the critical phenomena of the $XY$ model at low temperatures
depend sensitively on how the correlation function behaves, which
appears to lead completely different critical behaviors. It is also interesting
to note that the very same fact that all the vertices are closely 
connected by short path lengths yield very different behaviors, i.e., 
stimulating the mean-field nature for the Ising model, and
the enhancement of spinwave fluctuation to remove any ordered phase
in the $XY$ model.
This is not only an example that the effect of lattice geometry can 
solely change the criticality to a great extent, but also that the 
close connectedness does not guarantee the mean-field criticality.

We believe that the absence of an ordered phase
in the negatively curved system should apply in a broad range
of different cases, since it is based on the quadratic
approximation of the interaction which must be valid
for a variety of different interaction forms, including
Villain formulation of the $XY$ model.
Furthermore, even when the quadratic approximation fails, 
we believe the absence of an order should still be true,
since it appears to depend strongly on the topological  lattice
structure, rather than the detailed form of the interaction:
The exponentially decaying correlation function must be related
to the fact that the boundary increases exponentially with the
distance from any point. Since it is the basic geometric feature of a
negatively curved surface, not restricted to the heptagonal structure,
the physics will remain unaltered also in the continuum.

In summary, we have investigated the $XY$ model in heptagonal lattices 
in parallel to studies of 2D square lattices and the WS small-world networks.
Extensive Monte-Carlo simulations have been performed to compute
the Binder's cumulant, which has shown clearly that neither
the true long-range order nor the quasi-long-range KT order
is established at any finite temperatures. The vortex number density 
has been shown to be not the origin of the absence of an ordered phase. 
From the numerical calculation of the correlation function we
have found that the short path lengths facilitate the strong
spinwave fluctuation, which eventually removes any order, 
leading to the exponentially decaying
correlation function at any nonzero temperature.  We have
also discussed the recent works on the Ising model
in the same heptagonal lattice.

\acknowledgments
This work was supported by the Korea Research Foundation Grant funded by the
Korean Government(MOEHRD) (KRF-2005-005-J11903).

\end{document}